\documentclass[prc,showpacs,floatfix]{revtex4}
\usepackage{bm}
\usepackage{graphicx}
\newcommand{\beq}{\begin{equation}}
\newcommand{\eeq}{\end{equation}}
\newcommand{\beqa}{\begin{eqnarray}}
\newcommand{\eeqa}{\end{eqnarray}}

\begin{document}

\title{\bf Polarization observables in $\pi$-photoproduction on the
  deuteron\footnote{To appear in the Proceedings of {\it 4$^{th}$ Conference on
    Nuclear and Particle Physics, October 11-15, 2003, Fayoum, Egypt.}}}

\author{Eed M.\ Darwish\footnote{e-mail: eeddarwish@yahoo.com}}

\affiliation{Physics Department, Faculty of Science (Sohag), South Valley 
  University, Sohag 82524, Egypt}

\date{\today}

\begin{abstract}
A general analysis of polarization observables for the
$d(\gamma,\pi)NN$ reaction with polarized photons and/or oriented deuterons
is presented. The unpolarized differential cross section, linear photon
asymmetry, vector and tensor target asymmetries are predicted for forthcoming
experiments.
\end{abstract}

\maketitle
\section{Introduction}
Meson photo- and electroproduction on light nuclei is primarily
motivated by the following possibilities: (i) study of the elementary
neutron amplitude in the absence of a neutron target, (ii)
investigation of medium effects, i.e., possible changes of the
production operator in the presence of other nucleons, (iii) it
provides an interesting means to study nuclear structure, and (iv) 
it gives information on pion production on off-shell nucleon, as 
well as on the very important $\Delta N$-interaction in a nuclear 
medium. 

Polarization observables will give additional valuable information for 
checking the spin degrees of freedom of the elementary pion production 
amplitude of the neutron, provided, and this is very important, that 
one has under control all interfering interaction effects which prevent 
a simple extraction of this amplitude. As an illustration of these various 
aspects, incoherent single pion photoproduction on the deuteron in the
$\Delta$(1232)-resonance region is investigated with special emphasis on 
polarization observables~\cite{EedNew1,EedNew}. The importance of this process 
derives from the fact that the deuteron, being the simplest nuclear system, 
plays a similar fundamental role in nuclear physics as the hydrogen atom 
plays in atomic physics.

\section{Polarization observables}
The most general expression for all possible 
polarization observables in terms of the transition matrix elements is given 
by ~\cite{EedNew1,EedNew,Aren88}
\beqa
{\mathcal O} & = & \sum_{\alpha\alpha^{\prime}} \int_{0}^{q_{\rm max}}dq 
\int d\Omega_{p_{NN}}~\rho_s~ 
{\mathcal M}^{(t^{\prime}\mu^{\prime})~\star}_{s^{\prime}m^{\prime},
m_{\gamma}^{\prime}m_d^{\prime}} ~\vec{\Omega}_{s^{\prime}m^{\prime}sm}~
{\mathcal M}^{(t\mu)}_{sm,m_{\gamma}m_d}
~\rho^{\gamma}_{m_{\gamma}m_{\gamma}^{\prime}}~ \rho^{d}_{m_dm_{d}^{\prime}}\,,
\eeqa
where we have introduced as a shorthand for the quantum numbers 
$\alpha=(s, m,t,m_{\gamma},m_d)$ and $\alpha^{\prime}=(s^{\prime}, m^{\prime},t^{\prime},m_{\gamma}^{\prime},
m_d^{\prime})$. Furthermore, $m_{\gamma}$ denotes the photon
polarization, $m_{d}$ the spin projection of the deuteron, $s$ and $m$
total spin and projection of the two outgoing nucleons, respectively,
$t$ their total isospin, and $\mu$ the isospin projection of the pion. 
$\rho^{\gamma}_{m_{\gamma}m_{\gamma}^{\prime}}$ and
$\rho^{d}_{m_dm_{d}^{\prime}}$ denote the density matrices of initial
photon polarization and deuteron orientation, respectively, 
$\vec{\Omega}_{s^{\prime}m^{\prime}sm}$ is an operator associated with
the observable, which acts in the two-nucleon spin space. 
For details with respect to the density matrices we refer to~\cite{Aren88}. 
The transition $\mathcal{M}$-matrix elements of the $d(\gamma,\pi)NN$ 
reaction as well as the phase space factor $\rho_s$  are given 
in~\cite{Dar03}.

As in our previous work~\cite{EedNew}, the unpolarized differential cross 
section is given by
\beqa
\label{unpdcs}
\frac{d^3\sigma}{d\Omega_{\pi}dq} & = & \frac 16 ~\sum_{\alpha} \int d\Omega_{p_{NN}} 
  ~\rho_{s}~ |{\mathcal M}^{(t\mu)}_{sm m_{\gamma}m_d}|^{2}\,.
\eeqa

The photon asymmetry for linearly polarized photons is given by~\cite{EedNew1}
\beqa
\Sigma & = & \frac{2}{\mathcal A}~\Re e \sum_{smtm_{d}} 
\int_{0}^{q_{\rm max}}dq \int d\Omega_{p_{NN}}~ \rho_{s}~ 
{\mathcal M}^{(t\mu)}_{sm +1m_d}~ 
{\mathcal M}^{(t\mu)~\star}_{sm -1m_d}\,,
\label{FSigma}
\eeqa
with 
\beqa
\mathcal A & = & {\sum_{\alpha} \int_{0}^{q_{\rm max}}dq 
\int d\Omega_{p_{NN}}~\rho_{s} ~ 
|{\mathcal M}^{(t\mu)}_{sm m_{\gamma}m_d}|^{2}}\,.
\eeqa

The vector target asymmetry  $T_{11}$ is given by~\cite{EedNew1}
\beqa
T_{11} & = & \frac{\sqrt{6}}{\mathcal A} ~\Im m\sum_{smtm_{\gamma}} 
\int_{0}^{q_{\rm max}}dq \int d\Omega_{p_{NN}}~ \rho_{s}~ 
\Big[{\mathcal M}^{(t\mu)}_{smm_{\gamma}-1} 
- {\mathcal M}^{(t\mu)}_{smm_{\gamma}+1}\Big]
{\mathcal M}^{(t\mu)~\star}_{smm_{\gamma}0}\,. 
\label{FT11}
\eeqa

The tensor target asymmetries are given by~\cite{EedNew1}
\beqa
T_{20} & = &  \frac{1}{\sqrt{2}\mathcal A} ~
\sum_{smtm_{\gamma}}\int_{0}^{q_{\rm max}}dq \int d\Omega_{p_{NN}}~\rho_{s}~
\Big[|{\mathcal M}^{(t\mu)}_{smm_{\gamma}+1}|^2 
+ |{\mathcal M}^{(t\mu)}_{smm_{\gamma}-1}|^2 
- 2~ |{\mathcal M}^{(t\mu)}_{smm_{\gamma}0}|^2\Big] \,,
\label{FT20}
\eeqa
\beqa
T_{21} & = & \frac{\sqrt{6}}{\mathcal A} ~\Re e\sum_{smtm_{\gamma}} 
\int_{0}^{q_{\rm max}}dq \int d\Omega_{p_{NN}}~\rho_{s} ~
\Big[{\mathcal M}^{(t\mu)}_{smm_{\gamma}-1} 
- {\mathcal M}^{(t\mu)}_{smm_{\gamma}+1}\Big] 
{\mathcal M}^{(t\mu)~\star}_{smm_{\gamma}0} \,,
\label{FT21}
\eeqa
and
\beqa
T_{22} & = & \frac{2\sqrt{3}}{\mathcal A} ~\Re e\sum_{smtm_{\gamma}} 
\int_{0}^{q_{\rm max}}dq \int d\Omega_{p_{NN}} ~\rho_{s}~ 
{\mathcal M}^{(t\mu)}_{smm_{\gamma}-1} 
~{\mathcal M}^{(t\mu)~\star}_{smm_{\gamma}+1}\,. 
\label{FT22}
\eeqa

\section{Results and discussion}
For the evaluation of the above mentioned observables we have 
chosen the laboratory frame of the deuteron. As coordinate system a 
right-handed one is taken with $z$-axis along the momentum $\vec k$ of the 
incoming photon and $y$-axis along $\vec k\times\vec q$, where $\vec q$ is the 
pion momentum. As seen above, all observables are calculated by 
integrating over the pion momentum $q$, the polar angle $\theta_{p_{NN}}$ and 
the azimuthal angle $\phi_{p_{NN}}$ of the relative momentum $\vec p_{NN}$ of the two outgoing 
nucleons. These integrations are carried out numerically. The number of 
integration points was being increased until the accuracy of calculated 
observable becomes good to 1$\%$. 

The contribution to the pion 
production amplitude is evaluated by taking a realistic $NN$ 
potential model for the deuteron wave function. For our calculations we have 
used the wave function of the Paris potential~\cite{La+81}, which is in 
excellent agreement with $NN$ scattering data~\cite{Dar03th}. 
For the elementary pion photoproduction operator, we have taken the 
effective Lagrangian model of Schmidt {\it et al.}~\cite{ScA96}. This 
model had been constructed to give a realistic description of the 
$\Delta$(1232)-resonance region. It is given in an arbitrary frame of 
reference and allows a well defined off-shell continuation as required 
for studying pion production on nuclei. Therefore, this model for the 
elementary photoproduction amplitude is quite satisfactory for our purpose, 
namely to incorporate it into the reaction on the deuteron.

\begin{figure}[htb]
\begin{center}
\includegraphics[scale=.7]{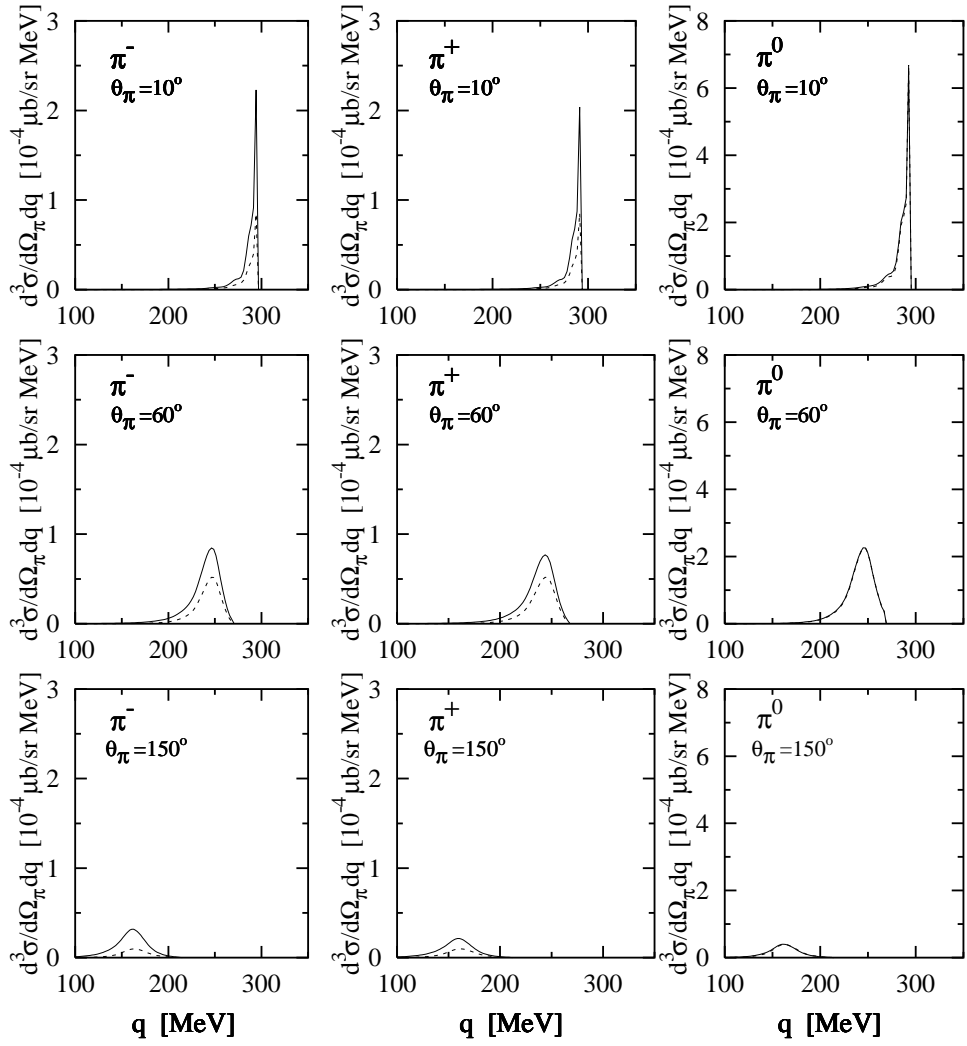}
\caption{The $\pi$-meson spectra in the $d(\gamma,\pi)NN$ reaction as a 
function of pion momentum $q$ at a photon energy of 330 
MeV for different values of pion angles $\theta_{\pi}$. The solid curves 
show the results of the full calculations while the dashed curves represent 
the results when only the $\Delta$(1232)-resonance is taken into account. 
The left, middle and right panels represent the results for 
$\gamma d\to\pi^-pp$, $\pi^+nn$ and $\pi^0np$, respectively.}  
\label{unpolcs1}
\end{center}
\end{figure}

Fig.~\ref{unpolcs1} shows our results for the $\pi$-meson spectra 
(\ref{unpdcs}) as a function of pion momentum $q$. One sees 
that when $q$ reaches its maximum, the absolute value of the relative momentum 
$p_{NN}$ of the two outgoing nucleons vanishes, and thus a narrow 
peak is appears in the forward pion angles. In the lower part of Fig.~\ref{unpolcs1} we see that 
the unpolarized differential cross section is small and the narrow peak which 
appears at forward angles is disappears. The same effect appears in the 
coherent process of charged pion photo- and electroproduction on the 
deuteron~\cite{Lag78,Kob87}, in deuteron electrodisintegration~\cite{Fab76} as 
well as in $\eta$-photoproduction~\cite{Fix97}. It is also clear that the 
maximum value of $q$ (when $p_{NN}\to 0$) decreases with increasing the 
pion angle. It changes from $\sim$ 300 MeV at forward angles to $\sim$ 200 MeV 
at backward ones. In principle, the experimental observation of this peak in 
the high $\pi$-momentum spectrum may serve as another evidence for the 
understanding of the $\pi$-meson spectra.

\begin{figure}[htb]
\begin{center}
\includegraphics[scale=.7]{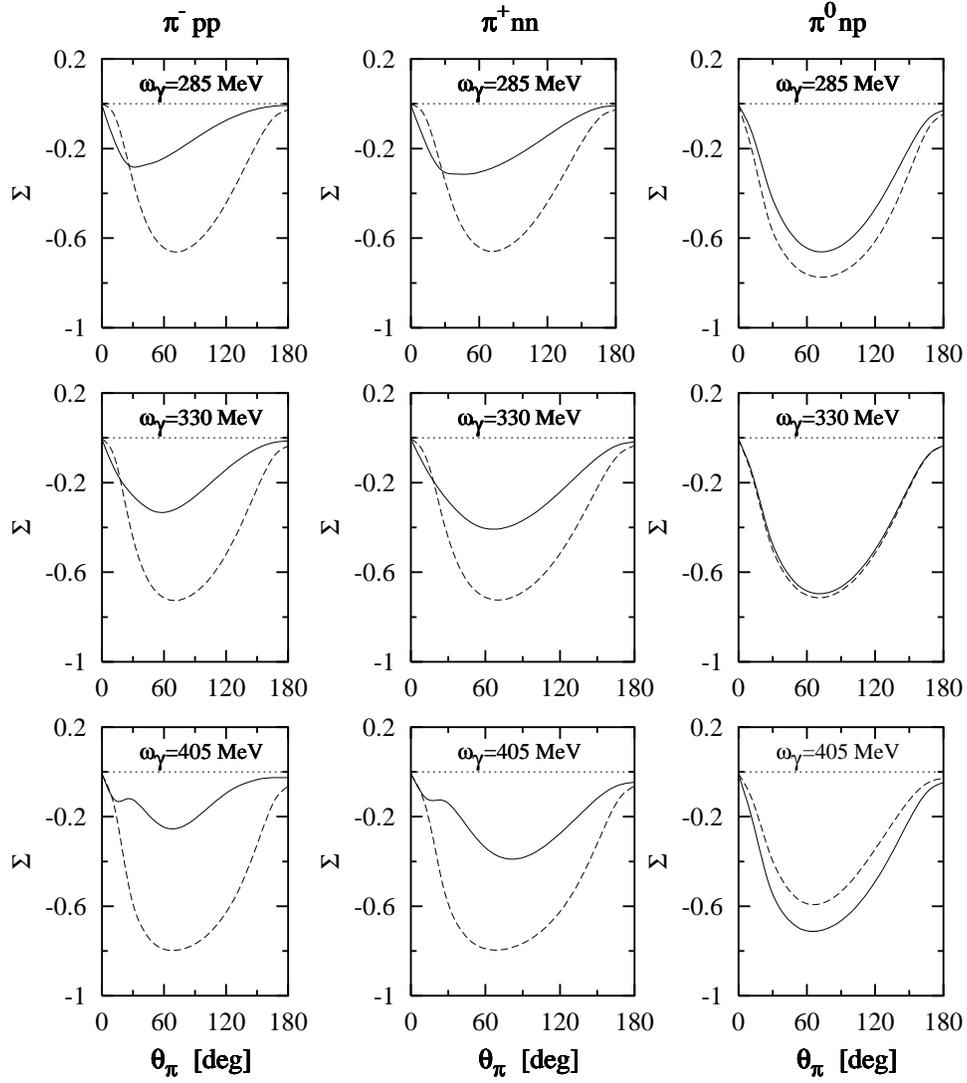}
\caption{Linear photon asymmetry of $d(\vec\gamma,\pi)NN$. 
  Notation as in Fig.~\ref{unpolcs1}.}
\label{phasym1}
\end{center}
\end{figure}

Our results for the photon asymmetry $\Sigma$ for 
linearly polarized photons (\ref{FSigma}) for all different charge 
states of the pion of $d(\vec\gamma,\pi)NN$ are plotted in 
Fig.~\ref{phasym1} at different photon lab-energies as a function of pion 
angle $\theta_{\pi}$ in the laboratory frame. First of all, we see that the
photon asymmetry has always a negative values at forward and backward
pion angles. One notes qualitatively a similar behaviour for charged pion channels whereas
a totally different behaviour is seen for the neutral pion channel.

For extreme forward and backward pion angles one sees, that the
effect of Born contributions is relatively small in comparison to the
results when $\theta_{\pi}$ changes from $60^{\circ}$ to $120^{\circ}$.
One notices also, that the contribution from Born terms are much important in
this region. In the energy range of 
the $\Delta$(1232)-resonance, one sees that the contribution from Born terms 
are important in the case of charged pion channels. For the neutral pion 
channel we see, that this contribution is very small at 330 MeV. 
For lower and higher energies, one sees 
again the sizeable effect from Born terms which arise mainly from the 
Kroll-Rudermann term. One 
sees also, that $\Sigma$ is sensitive to the energy of the incoming photon. 
Finally, we observe that the interference of the Born terms
with the $\Delta$(1232)-resonance contribution causes considerable changes in
the linear photon asymmetry. Experimental measurements will give us more
valuable information on this asymmetry.

\begin{figure}[htb]
\begin{center}
\includegraphics[scale=.7]{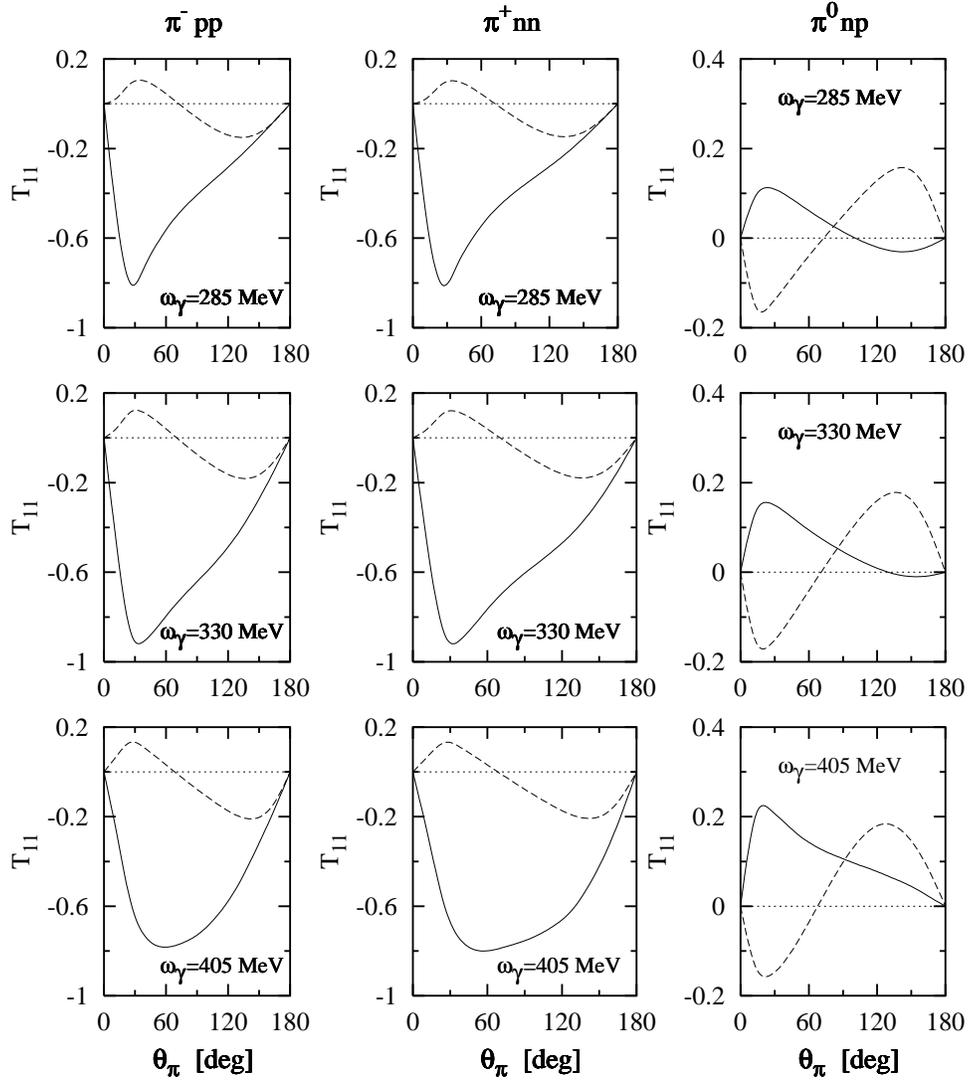}
\caption{Vector target asymmetry of $\vec d(\gamma,\pi)NN$. 
  Notation as in Fig.~\ref{unpolcs1}.}
\label{vtasym1}
\end{center}
\end{figure}

Our results for the vector target 
asymmetry $T_{11}$ (\ref{FT11}) are depicted in Fig.~\ref{vtasym1} for $\gamma \vec d\to\pi^-pp$ (left panels), 
$\pi^+nn$ (middle panels), and $\pi^0np$ (right panels), respectively. 
The asymmetry $T_{11}$ clearly differs in size between charged and neutral pion
photoproduction channels, being even opposite in phase. For charged pion
photoproduction reactions we see, that the vector target asymmetry has always a negative
values which mainly come from the Born terms. A small positive contribution
from the $\Delta$-resonance is found only at pion forward angles. At backward
angles, the negative values for $T_{11}$ come from an interference of the Born
terms with the $\Delta$(1232)-resonance contribution. For all energies one
observes at forward angle the strongest effect of the Born terms.

With respect to the neutral pion photoproduction channel, we see that the 
vector target 
asymmetry is always positive. For energies below the $\Delta$-resonance,
a very small negative value is found at extreme backward pion angles while a
relatively large positive value at forward angles is found. 
We see also, that $T_{11}$ is sensitive to the Born terms. 
The same effect was found by Blaazer {\it et al.}~\cite{Bla94} and Wilhelm and
Arenh\"ovel~\cite{Wil95} for the coherent pion photoproduction reaction
on the deuteron. The reason is that $T_{11}$ depends on the relative phase
of the matrix elements as can be seen from (\ref{FT11}).
It would vanish for a constant overall phase of the $t$-matrix, a case which
is approximately realized if only the $\Delta$(1232)-amplitude is considered.
Finally, we notice that $T_{11}$ is vanishes at $\theta_{\pi}=0$ and $\theta_{\pi}=\pi$
which is not the case for the linear photon asymmetry.

\begin{figure}[htb]
\begin{center}
\includegraphics[scale=.7]{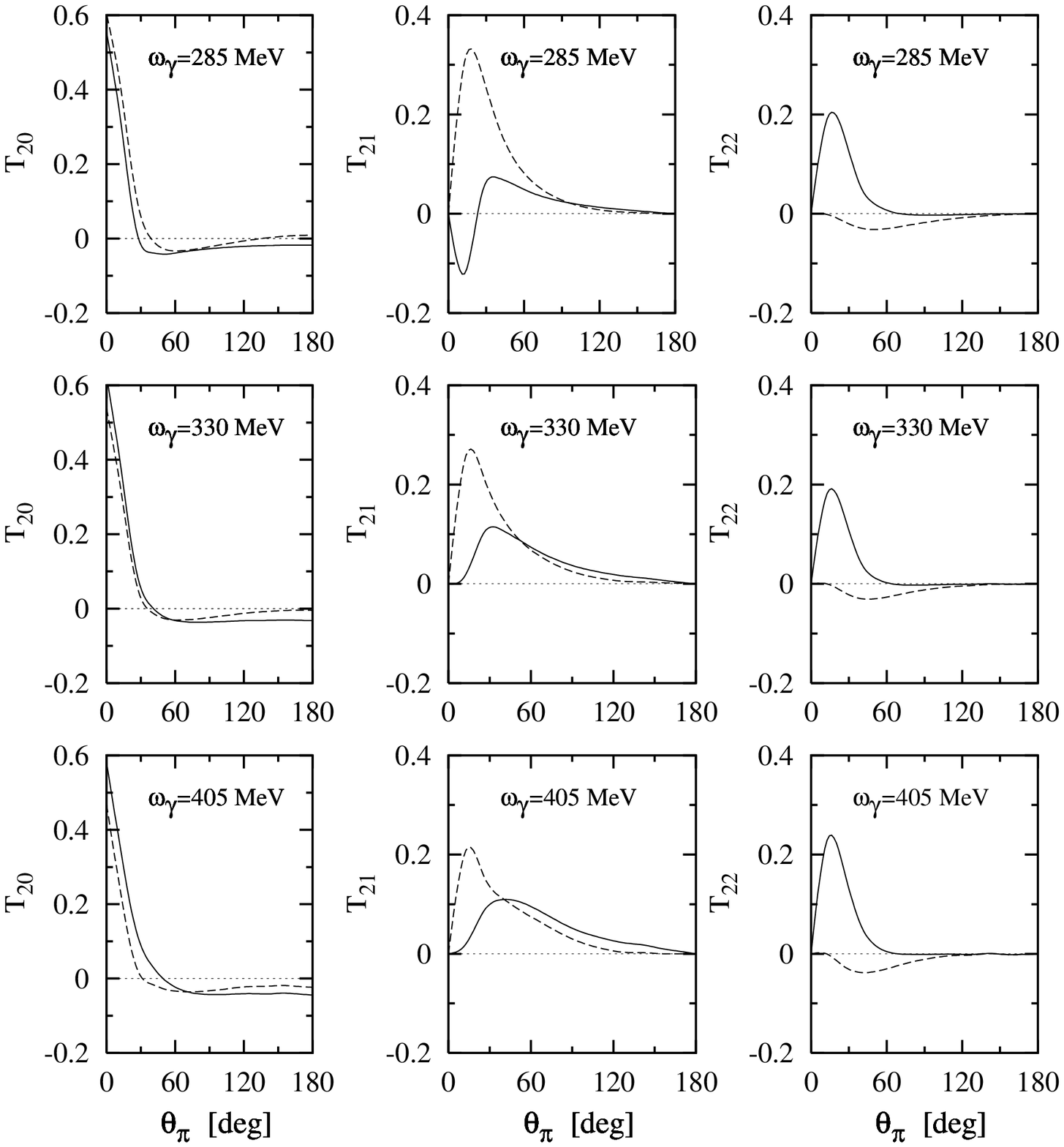}
\caption{Tensor target asymmetries of $\vec d(\gamma,\pi^-)pp$. 
  Notation as in Fig.~\ref{unpolcs1}.}
\label{ttasym201}
\end{center}
\end{figure}

\begin{figure}[htb]
\begin{center}
\includegraphics[scale=.7]{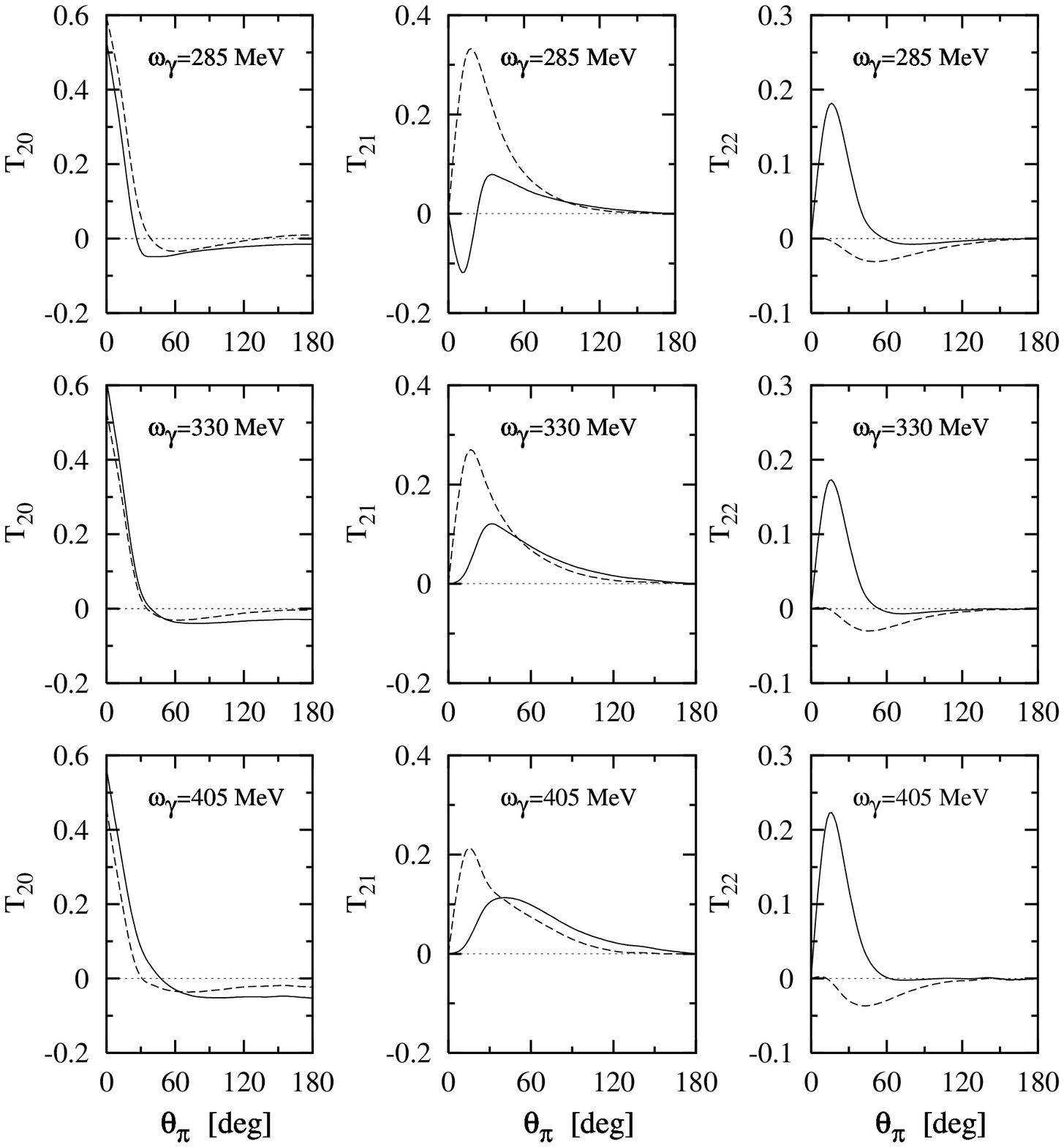}
\caption{Tensor target asymmetries of $\vec d(\gamma,\pi^+)nn$. 
  Notation as in Fig.~\ref{unpolcs1}.}
\label{ttasym211}
\end{center}
\end{figure}

\begin{figure}[htb]
\begin{center}
\includegraphics[scale=.7]{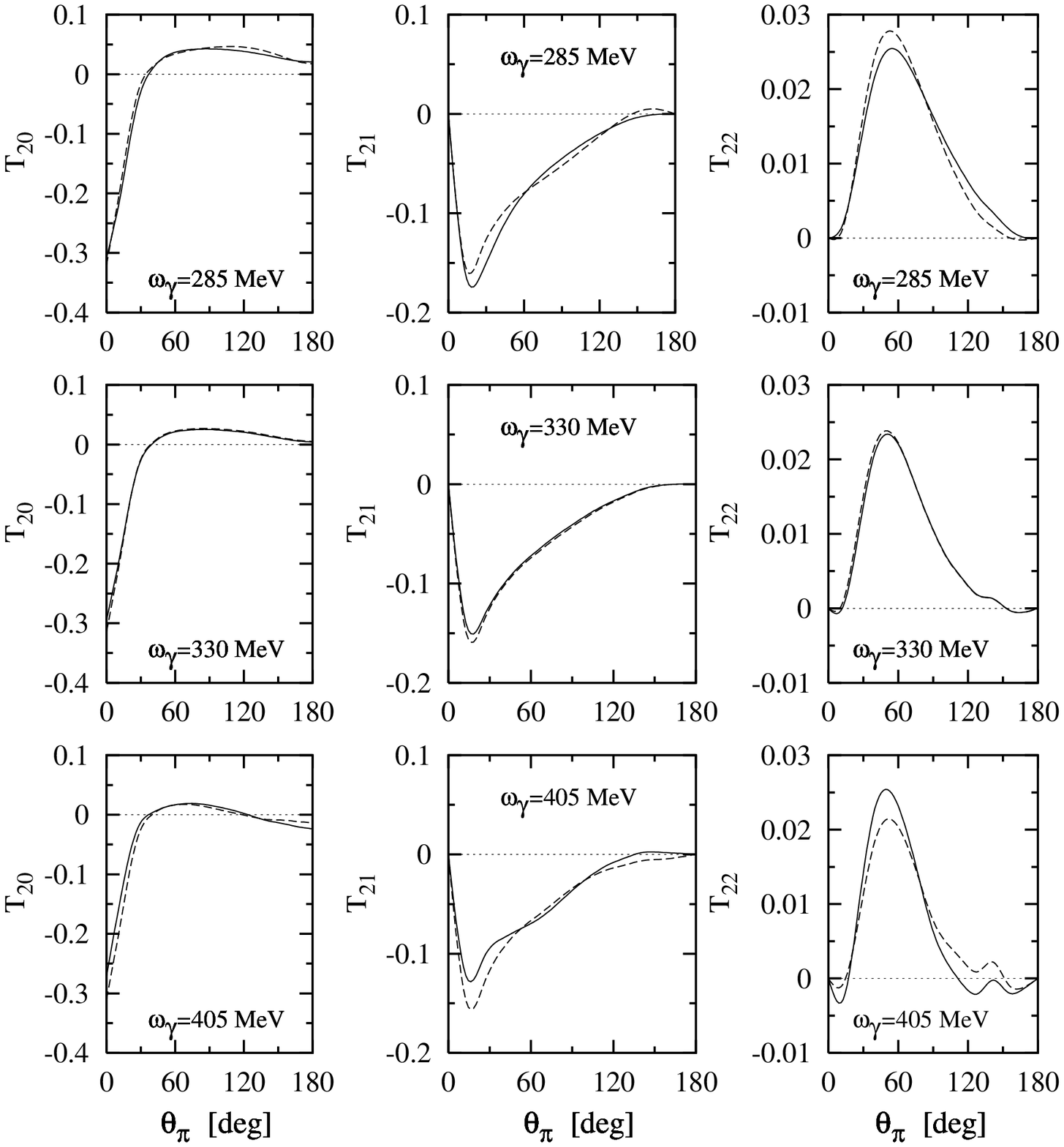}
\caption{Tensor target asymmetries of $\vec d(\gamma,\pi^0)np$. 
  Notation as in Fig.~\ref{unpolcs1}.}
\label{ttasym221}
\end{center}
\end{figure}

Let us discuss now the results of the tensor target asymmetries
$T_{20}$ (\ref{FT20}), $T_{21}$ (\ref{FT21}), and $T_{22}$ (\ref{FT22}) 
as shown in Figs.~\ref{ttasym201},
\ref{ttasym211}, and \ref{ttasym221} for $\gamma \vec d\to\pi^-pp$,
$\pi^+nn$, and $\pi^0np$, respectively. We start from the tensor asymmetry
$T_{20}$ which is plotted in the left panels. For $\gamma d\to\pi NN$ at 
forward and backward pion angles, the asymmetry $T_{20}$
allows one to draw specific conclusions about details of the reaction
mechanism. In comparison to the results for photon and vector target asymmetries
we found here, that the contribution from the Born terms is very small both for charged
and neutral pion production channels. It is also noticeable, that for charged channels
the asymmetry $T_{20}$ has a relatively large positive values at pion forward angles while a 
small negative ones at backward angles are found.
For the neutral pion production channel we see, that $T_{20}$ has a negative values
at forward angles and a positive ones at backward angles. Only for energies above the $\Delta$-resonance we note, that
it has a small negative values at extreme backward angles.

The tensor target asymmetry $T_{21}$ of $\gamma \vec d\to\pi^-pp$,
$\pi^+nn$, and $\pi^0np$ is plotted in the middle panels of
Figs.~\ref{ttasym201}, \ref{ttasym211}, and \ref{ttasym221}, respectively. It 
is clear that
$T_{21}$ differs in size between charged and neutral pion production channels.
One notices, that for charged pion channels $T_{21}$ asymmetry is sensitive to the Born terms, in
particular at forward pion angles. In the case of $\pi^0$ channel one sees,
that the contribution of the Born terms is much less important at all energies.
In comparison to the results for photon and vector target asymmetries
we found also here, that the contribution from the Born terms is small both
for charged and neutral pion production channels. It is also noticeable,
that in the case of charged pion channels the asymmetry $T_{21}$ has a
relatively large positive values at pion forward angles.
For the neutral pion channel we see, that $T_{21}$ has a negative values
at forward angles. Furthermore, as in the case of vector target asymmetry, we
found that $T_{21}$ is vanishes at $\theta_{\pi}=0$ and $\theta_{\pi}=\pi$.

In the right panels of Figs.~\ref{ttasym201}, \ref{ttasym211}, and \ref{ttasym221} we depict our
results for the tensor target asymmetry $T_{22}$ for the reactions
$\gamma \vec d\to\pi^-pp$, $\pi^+nn$, and $\pi^0np$, respectively.
One readily notes the importance of Born terms at extreme forward pion angles. Like
the results of the $T_{20}$ and $T_{21}$ asymmetries, the $T_{22}$ asymmetry
is sensitive to the values of pion angle $\theta_{\pi}$.
At $\theta_{\pi}=60^{\circ}$ we see, that the Born terms are
important for $\pi^0$ production channel while these terms are very important
for charged pion channels at extreme forward angles. Moreover, we found that
$T_{22}$ is also vanishes at $\theta_{\pi}=0$ and $\theta_{\pi}=\pi$.

\section{Conclusion}
I would like to conclude that the results presented here for polarization 
observables in the $d(\gamma,\pi)NN$ reaction in the $\Delta$(1232)-resonance 
region can be used as a basis for the simulation of the behaviour of 
polarization observables and for an optimal planning of new polarization 
experiments of this reaction. It would be very interesting to examine our 
predictions experimentally.

\section*{Acknowledgments}
I would like to express my gratitude to the organizing committee and the 
chairman of the NUPPAC'03 Conference for the warm hospitality and providing 
the pleasant environment for useful discussions. Very useful discussions with 
Professor H.\ Arenh\"ovel are gratefully acknowledge. 


\newpage
\centerline{FIGURES}

\end{document}